\documentclass[aps, twocolumn, showpacs, amsmath]{revtex4}
\usepackage{graphicx,epsfig}

\DeclareGraphicsExtensions{. jpg,. pdf, . mps, .png, .eps, . ps, . EPS}

\usepackage{amsmath}
\usepackage{color}
\begin{document}
\newcommand{\beq}{\begin{equation}}
\newcommand{\eeq}{\end{equation}}
\newcommand{\bea}{\begin{eqnarray}}
\newcommand{\eea}{\end{eqnarray}}
\newcommand{\gt}{\tilde{g}}
\newcommand{\mt}{\tilde{\mu}}
\newcommand{\et}{\tilde{\varepsilon}}
\newcommand{\ct}{\tilde{C}}
\newcommand{\bt}{\tilde{\beta}}

\newcommand{\avg}[1]{\langle{#1}\rangle}
\newcommand{\Avg}[1]{\left\langle{#1}\right\rangle}
\newcommand{\cor}[1]{\textcolor{red}{#1}}

\title{Network  Controllability Is Determined by the Density of Low In-Degree \\and Out-Degree Nodes }
\author{Giulia Menichetti}
\affiliation{Department of Physics and Astronomy and INFN Sez. Bologna, Bologna University, Viale B. Pichat 6/2 40127 Bologna, Italy}
\author{Luca Dall'Asta}
\affiliation{Department of Applied Science and Technology – DISAT, Politecnico di Torino, Corso Duca degli Abruzzi 24, 10129 Torino, Italy}
\affiliation{Collegio Carlo Alberto, Via Real Collegio 30, 10024 Moncalieri, Italy}
\author{Ginestra Bianconi}
\affiliation{School of Mathematical Sciences, Queen Mary University of London, London E1 4NS, United Kingdom}
\begin{abstract}
The problem of controllability of the dynamical state of a network is central in network theory and has   wide applications ranging from network medicine  to financial markets. The driver nodes of the network are the nodes that can bring the network to the desired dynamical state if an external signal is applied to them. Using the framework of structural controllability, here  we show that the density of nodes with  in-degree and  out-degree equal to $0$,  $1$ and $2$  determines  the number of driver nodes of random networks. Moreover we show that  networks with minimum in-degree  and out-degree greater than 2, are always fully controllable by an infinitesimal fraction of driver nodes, regardless on the other properties of the degree distribution.  Finally, based on these results,  we propose an algorithm to improve the controllability of networks.
\end{abstract}
\maketitle

The controllability of a network \cite{slotine_book,con_pinning,con_sorrentino,con_boccaletti,Lin74,Liu,correlations,con_grebogi,con_lai,con_vicsek} is a fundamental problem with wide applications ranging from medicine and drug discovery \cite{drug_discovery}, to  the characterization of dynamical processes in the brain \cite{Bullmore, Bonifazi,brain}, or the evaluation of risk in financial markets \cite{control_Caldarelli}.
While the interplay between the structure of the network \cite{RMP,Newman_rev,Boccaletti2006,Caldarelli} and the dynamical processes defined on them has been an active subject of complex network research for more than ten years \cite{crit,Dynamics}, only recently the rich interplay between the controllability of a network  and its structure has started to be investigated.
A pivotal role in this respect has been played by a paper by Liu et al. \cite{Liu}, in which the problem of finding the minimal set of driver
nodes necessary to control a network was mapped into a maximum matching problem. Using a well established statistical mechanics approach \cite{Cavity, Zecchina,Weigt,Mezard,Lenka,Altarelli},
Liu et al. \cite{Liu} characterize in detail the set of driver nodes for real networks and for ensembles of networks with given in-degree and out-degree distribution. By analyzing scale-free networks with minimum in-degree and minimum out-degree  equal to 1 they have found that the smaller is the power-law exponent $\gamma$ of the degree distribution, the larger is the fraction of driver nodes in the network. 
This result has prompted the authors of \cite{Liu} to say that the higher is the heterogeneity of the degree distribution, the less controllable is the network. Later, different papers  have addressed questions related to controllability of networks with similar tools \cite{correlations,bimodality}.

In this Letter we consider the network controllability  and its mapping to the maximum matching problem, exploring the role of low in-degree and low out-degree nodes in the network.
We show  that by changing the fraction of nodes with in-degree and out-degree less than 3, the number of driver nodes of a network can change in a dramatic way. In particular if the minimum in-degree and the minimum out-degree of a network are both greater than 2 then any network, independently on the level of heterogeneity of the degree distribution,  is fully controllable by an infinitesimal fraction of nodes.
Therefore we show that the heterogeneity of the network is not the only element determining the number of driver nodes in the network and that this number is very sensible on the fraction of  low in-degree low  out-degree nodes of the network. 
This result allows us to propose a method to improve the controllability of networks by decreasing the density of nodes with in-degree and out-degree less than $3$, adding links to the network.

{\it The structural controllability of a network.} 
Given a graph $G = (V,E)$ of $N$ nodes, we consider a continuous-time linear dynamical system  
\bea
\frac{d{\bf x}(t)}{dt}=A{\bf x}+B{\bf u},
\eea
in which the vector ${\bf x}(t)$, of elements $x_i(t)$ with $i=1,2,\ldots, N$, represents the dynamical state of the network, $A$ is $N\times N$ (asymmetric) matrix describing the directed weighted interactions within the network, and     $B$ is a $N\times M$ matrix describing the interaction between the nodes of the graph and $M\leq N$ external signals, indicated by the vector ${\bf u}(t)$ of elements $u_{\alpha}$ and $\alpha=1,2\ldots M$. For any given realization of $A$ and $B$,  the dynamical system is controllable if it satisfies Kalman's controllability rank condition, i.e. the matrix $C=(B,AB, A^2B, \dots, A^{N-1}B)$ is full rank. In addition to the fact that the verification of Kalman's condition can be computationally very demanding for large systems, in most real systems the notion of exact controllability is unusable since the entries of $A$ and $B$ are not perfectly known. As an alternative, if we assume that the non-zero matrix elements of $A$ and $B$ are free parameters, we can consider the concept of {\em structural controllability} \cite{Lin74}. The system is structurally controllable if for any choice of the free parameters in $A$ and $B$, except for a variety of zero Lebesgue measure in the parameter space,  $C$ is full rank \cite{Lin74}. 
Since structural controllability only distinguishes between zero and non-zero entries of the matrices $A$ and $B$, a given directed network is structurally controllable if it is possible to determine the input nodes (i.e. the position of the non-zero entries of the matrix $B$) in a way to control the dynamics described by any realization of the matrix $A$ with the same non-zero elements, except for atypical realizations of zero measure.
In practice, a network can be structurally controlled by identifying a minimum number of {\em driver nodes}, that are controlled nodes which do not share input vertices. 
In their seminal paper \cite{Liu}, Liu and coworkers showed that this control theoretic problem can be reduced to a well-known optimization problem: their Minimum Input Theorem states that the minimum set of driver nodes that guarantees the full structural controllability of a network is the set of unmatched nodes in a maximum  matching of the same directed network.

{\it The maximum matching problem.}\ A  matching $M$ of a directed  graph is a set of directed edges without common start or end vertices, and it is maximum when it contains the maximum possible number of edges. 
The problem of finding a maximum matching of a directed graph can be cast on a statistical mechanics problem, by introducing variables $s_{ij} \in \{1,0\}$ on each directed link from node $i$ to node $j$, indicating whether the directed link is in $M$ ($s_{ij} = 1$) or not ($s_{ij} = 0$). The configurations of variables $\{s_{ij}\}$ have to satisfy the following matching condition, 
\bea
\sum_{j\in \partial_{+}i}s_{ij}\leq 1,\ \ \ \ \ \ \sum_{j\in \partial_{-}i}s_{ji}\leq 1,
\eea 
where $\partial_- i$ indicates the set of  nodes $j$ that point to node $i$ in the directed network,  and $\partial_+ i$ indicates the set of nodes $j$ that  are pointed by node $i$. Moreover the variables $\{s_{ij}\}$  should minimize the energy function 
\bea
E=2\sum_{i=1}^N \left( 1-\sum_{j\in \partial_{-}i}s_{ji} \right). 
\label{energy}
\eea

Note that a vertex is matched if it is the endpoint of one of the edges in the matching, otherwise the vertex is {\em unmatched}. It follows that $E = 2 N_D$,  
where $N_D$ is the number of unmatched nodes in the network, and this number also determines the minimum number of driver nodes required to fully control the network.
Following Refs.\cite{Lenka,Liu}, we use the cavity method in the zero-temperature limit, to study the statistical properties of maximum matchings on directed random graphs for which the locally-tree-like approximation holds. 
Under the decorrelation (replica-symmetric) assumption, the energy of a maximum matching can be written in terms of the cavity fields (or messages) $h_{i\to j}$ or $\hat{h}_{i\to j}$ sent from a node $i$ to the linked node $j$. The fields are sent in the same direction $h_{i\to j}$ or in the opposite direction $\hat{h}_{i\to j}$ of the links and indicate the following messages \cite{Lenka}: $h_{i\to j}=\hat{h}_{i\to j}=1$  indicates {\em match me}, $h_{i\to j}=\hat{h}_{i\to j}=-1$ indicates {\em do not match me}, finally  $h_{i\to j}=\hat{h}_{i\to j}=0$ indicates {\em do what you want}. In fact the energy $E$ follows (see Supplemental Material (SM) \cite{supplemental} for details)
\bea
\nonumber E & = & - \sum_{i=1}^N \max\left[-1, \max_{k\in\partial_{+}i} \hat{h}_{k \to i}\right] - \sum_{i=1}^N \max\left[-1, \max_{k\in\partial_{-}i} h_{k \to i}\right] \\
  & & + \sum_{<i,j>}\max\left[0, h_{i \to j}+\hat{h}_{j \to i}\right]
\eea
in which for each directed link $(i,j)$ the cavity fields $\{h_{i \to j}, \hat{h}_{i \to j}\}$ satisfy the following zero-temperature version of the Belief Propagation (BP) equations, also known as Max-Sum  (MS) equations,  
\begin{subequations}
\bea
\label{MSa} h_{i \to j}&=&-\max\left[-1, \max_{k\in\partial_{+}i\setminus j} \hat{h}_{k \to i}\right], \\
\label{MSb} \hat{h}_{i \to j}&=&-\max\left[-1, \max_{k\in\partial_{-}i\setminus j} h_{k \to i}\right],
\eea
\end{subequations}
with the assumption that the maximum over an empty set is equal to $-1$. 
In the infinite size limit, the MS equations are closed for cavity fields with support  on $\{-1,0,1\}$ \cite{Lenka,Liu, Altarelli}.  These equation can be solved by iteration using  the BP/MS algorithm. 
\\
{\it Sufficient condition for the full controllability of networks.}
Let us now show  that for any network topology if the in-degree and the out-degree of the network is greater than 2 the fraction of driver nodes is zero. First we  observe that  the configuration in which all fields are zero , i.e. $h_{i \to j}=\hat{h}_{i\to j}=0$,  is an allowed solution of the Eqs. $(\ref{MSa})-(\ref{MSb})$  as soon as the minimum  in-degree and minimum out-degree equal to 1. In fact if a node has  in-degree 1  this link must be matched, and a similar situation occurs for the nodes with out-degree 1, generating a set of hard constraints incompatible with the configuration in which all the fields are zero, while if the minimum in-degree or out-degree of the network is greater than 1, all the nodes can be matched in a variety of ways therefore all the fields can be equal to zero. This solution corresponds to a fraction of driver nodes $n_D=0$ if the minimum in-degree and the minimum out-degree are greater than 1. This solution is also stable if, when we change a single field from zero to a value different from zero,  the perturbation does not propagate in the network. Suppose that $\hat{h}_{k\to i}$ is changed, say, from $0$ to $1$, meaning that the message is {\em match me}, then all the nodes $j\in \partial_{+}i$ neighbor of $i$ and different from $k$ receive a message {\em do not match me}. But if all the nodes $j$ have more than 2 incoming links, also if the link $(j,k)$ is not matched they can still send to their incoming neighbors the messages {\em do what you want} since there are different ways in which the matching can be achieved and they do not have to impose  to any of their other links to be matched.  Therefore the perturbation does not propagate in the network.
  A similar argument holds for a change of the field $h_{k\to i}$ to 1 which does not propagate if the out-degree of the network is greater than 2. This stability argument shows that for every tree-like network for which the BP/MS equations are valid, if the in-degree and the out-degree of the network is greater than 2 then the density of driver nodes is $n_D=0$. Note that this a sufficient condition for the stability of the $n_D=0$ solution but more stringent conditions are discussed in the following for  networks with given degree distribution.

\begin{figure}
\begin{center}
\hspace*{-15mm}\includegraphics[width=3.0in]{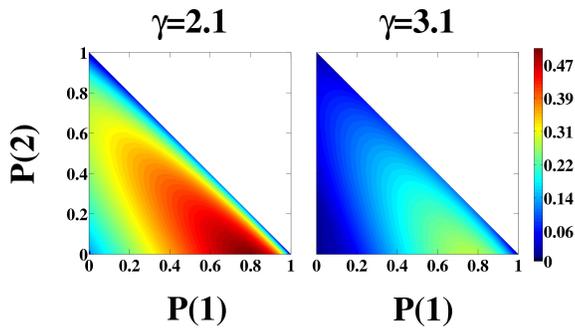}
\end{center}
\caption{Heat map representing the density of driver nodes $n_D$  as a function of the parameters $P(1)$ and $P(2)$ for networks of $N=10^6$ nodes with degree distribution given by Eq. \eqref{pkinout} and $\gamma=2.1$ (left), $3.1$ (right). The density $n_D$ is obtained by numerically solving the BP/MS equations for an ensemble of networks with given degree distribution. The region in which $P(1)+P(2)>1$ is non-physical.} 
\label{fig_tri}
\end{figure}
  
{\it Conditions for the full controllability of random networks.}
In the following we focus on ensembles of random networks with given in-degree and out-degree distribution $P^{in}(k)$ and $P^{out}(k)$. In this case (see SM \cite{supplemental}), it is possible to write the BP/MS equations and the energy  in terms of the probabilities  $w_i\in [0,1]$ and $\hat{w}_i\in [0,1]$ with $i=1,2,3$ that the cavity fields $h_{i\to j}$ and $\hat{h}_{i\to j}$ are respectively given by $\{1,-1,0\}$. 
\begin{figure}
\begin{center}
\includegraphics[width=2.0in]{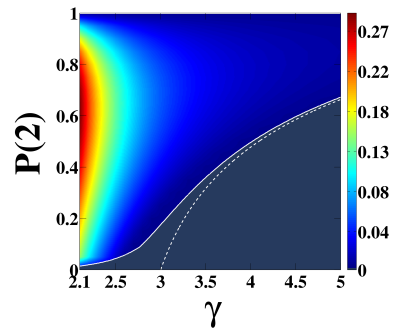} 
\end{center}
\caption{Phase diagram of the density of driver nodes $n_D$  as a function of the parameters $\gamma$ and $P(2)$ for networks of $N=10^6$ nodes with degree distribution given by Eq. $(\ref{pkinout})$ and $P(1)=0$. The density $n_D$ is obtained by numerically solving the BP/MS equations for an ensemble of networks with given degree distribution. The solid lines indicate the stability lines for $N=10^6$, the dotted lines indicate the stability lines in the limit $N\to \infty$. } 
\label{fig:phasediagram}
\end{figure}
From the BP/MS equations of the matching problem on random networks with given degree distribution, we found that the solution $n_D=0$ is allowed  if and only if $P^{in/out}(0)=P^{in/out}(1)=0$.  The replica-symmetric cavity equations are supposed to give the correct solution to the maximum matching problem if no instabilities take place. 
By analysing the stability condition of the BP/MS equations \cite{supplemental}, we find that the stability conditions for this solution in an ensemble of networks with given in-degree and out-degree sequence, are  
\bea
\hspace*{-3mm}P^{out}(2)<\frac{{\avg{k}_{in}}^2}{2\avg{k(k-1)}_{in}}, \ P^{in}(2)<\frac{{\avg{k}_{in}}^2}{2\avg{k(k-1)}_{out}}.
\label{stability}
\eea
In particular when the minimum in-degree and the minimum out-degree of scale-free networks are both greater than $2$, i.e. $P^{in/out}(0)=P^{in/out}(1)=P^{in/out}(2)=0$, the fraction of driver nodes is zero in the thermodynamic limit, for any choice of the degree distribution with this property. By changing the minimum in-degree and minimum out-degree of the network the number of driver nodes can change dramatically, independently of the tail of the degree distribution and the level of degree heterogeneity.

In order to use the above calculation to estimate the role of low-degree nodes on the fate of the zero-energy solution in finite networks, we consider uncorrelated random graphs with the following power-law degree distribution 
\bea
P^{in}(k)=P^{out}(k)=\left\{\begin{array}{lcc}
P(1) &\mbox{if}&{k=1}\\
P(2)&\mbox{if}& k=2 \\
Ck^{-\gamma} &\mbox{if} &k\in [3,K]\end{array}\right.
\label{pkinout}
\eea
with $C$ a constant determined by normalization and maximum degree $K=\min(\sqrt{N},\left\{[1-P(1)-P(2)]N\right\}^{1/(\gamma-1)})$ for $\gamma>2$ and $K=\min({N}^{1/\gamma},\left\{[1-P(1)-P(2)]N\right\}^{1/(\gamma-1)})$  for $\gamma\in (1,2]$, that is the minimum between the structural cutoff \cite{Marian_cutoff,GM} of the network and the natural cutoff of the degree distribution.
These networks can be generated numerically using the   configuration model. As long as $P(1)=P(2)=0$, the density of driver nodes goes to zero  ($n_D \to 0$) for any exponent $\gamma>1$. More generally, the density $n_D$ of driver nodes changes dramatically as a function of $P(1)$ and $P(2)$ as shown by the heat map in Fig.~\ref{fig_tri} for $\gamma =2.1, 3.1$. 
Moreover, in Fig.~\ref{fig:phasediagram},  we plot the phase diagram for $P(1)=0$ indicating the region where the solution $n_D=0$ is stable both for a finite network of $N=10^6$ nodes (white solid line) and for $N\to \infty$ (white dotted line).
Note that, for $\gamma \in (2,3]$, stability line converges quite slowly to zero in the infinite size limit. 

\begin{figure}[t!]
\begin{center}
\includegraphics[width=0.78\columnwidth]{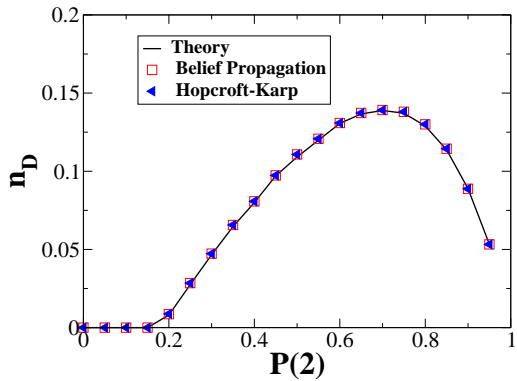} 
\end{center}
\caption{Density of driver nodes $n_D$ as a function of $P(2)$ for in-degree and out-degree distributions as in Eq. \eqref{pkinout} with $P(1)=0$ and $\gamma=2.3$. The fraction of driver nodes computed with the BP/MS algorithm on a network of $N=10^4$ nodes (averaged over $50$ network realizations) is compared with the exact results obtained using the Hopcroft-Karp algorithm for maximum matching \cite{Hopcroft} and with the theoretical expectation for the density $n_D$ in an ensemble of random networks with the same degree distribution.} 
\label{fig:3}
\end{figure}

A confirmation of the validity of this scenario is reported in Fig.~\ref{fig:3} from a direct comparison of the theoretical results in the ensemble of networks with given degree distribution, with those obtained by the BP algorithm or by computing explicitly the maximum matching using the Hopcroft-Karp algorithm \cite{Hopcroft} finding very good agreement. Fig.~\ref{fig:3} also shows that $n_D$ vanishes by decreasing $P(2)$. From our numerical results (reported in the SM \cite{supplemental}), in the region in which the solution $n_D=0$ is stable and we are far from the stability transition, both algorithms give a zero number of driver nodes $N_D=0$, meaning that all the nodes are matched, and therefore a single external input can be used to control the network.
 
{\it Improving the controllability of a network.}
These results suggest a simple and very effective way to improve the controllability of a network, by decreasing the fraction of nodes with in-degree and out-degree equal to $0$, $1$ and $2$. Starting from a  network with given degree distribution, we first add links starting from any node of out-degree equal to $0$ (if present in the network)  and randomly attached to any other node of the network, or starting from any random node of the network and ending to nodes of in-degree $0$. When there are no more nodes with in-degree or out-degree equal to $0$, we repeat the process of random addition of links to nodes with in-degree or out-degree equal to $1$ and $2$. At the end of the process the minimum in-degree of the network and the minimum out-degree is equal to $3$.

\begin{figure}[t]
\begin{center}
\includegraphics[width=0.78\columnwidth]{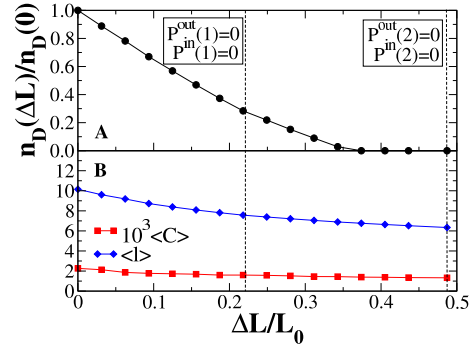} 
\end{center}
\caption{Fraction of driver nodes $n_D(\Delta L)/n_D(0)$ (panel A),  average clustering coefficient $\avg{C}$ and average distance $\avg{l}$ (panel B) of the network as a function of the fraction of added links to low degree nodes. The results are obtained from the BP/MS algorithm. The initial network is  a power-law network with in-degree distribution equal to the out-degree distribution,  $N=10^4$ nodes, and  power-law exponent $\gamma=2.3$. The symbol $\Delta L$ indicates the number of added links to the network, whereas $L_0$ indicates the initial number of links of the network.  } 
\label{fig:4}
\end{figure}
Fig.~\ref{fig:4}A shows the reduction in the fraction of driver nodes $n_D(\Delta L)$ compared to the original one $n_D(0)$ due to the addition of a fraction $\Delta L/L_0$ of directed links to a network with pure power-law degree distribution and structural cutoff. It is clear that by lowering the ratio of low in-degree and low out-degree nodes it is possible to reach full controllability of the network. However this can be costly, since for a given network the number of links that need to be added can be a significant fraction of the initial number of links. Nevertheless, by means of this link-addition process, the number of driver nodes decreases steadily and, for example, in the case considered in Fig.~\ref{fig:4} the number of driver nodes is decreased by $50\%$ just by adding a $12\%$ of links. Finally we have measured how other properties of the network change during this procedure, observing that the clustering coefficient does not change significantly while the average distance decreases. Note that this procedure can also be applied to networks with other degree distributions as Poisson networks (see SM \cite{supplemental}).

{\it Conclusions.}
We have shown that the structural controllability of a network depends strongly on the fraction of low in-degree and low out-degree nodes. For any uncorrelated directed network with given in-degree and out-degree distribution, the minimum fraction of driver nodes is zero, i.e. $n_D=0$, if the in-degrees and the out-degrees of all nodes are both greater than 2. For the relevant class of networks with power-law degree distribution, the number of driver nodes can change dramatically by changing the fraction of nodes with in-degree and out-degree equal to $1$ or $2$.
Finally we have proposed a strategy to improve the structural controllability of networks by adding links to low degree nodes. Since studying the controllability of real networks is essential for drug design, business applications and to study the stability of financial markets, we believe that our results will improve the understanding of controllability in such systems.
\appendix

\section{ The BP approach to the maximum matching problem}
\subsection{The maximum matching problem}
 The maximum matching problem  can be treated by statistical mechanics techniques 
\cite{Liu,Lenka,Altarelli,Cavity, Zecchina, Weigt,Mezard} such as the cavity method also known as Belief Propagation (BP). The problem on a directed network, is defined as follows \cite{Liu}.
On each link starting from  node $i$ and ending to node $j$ we  consider the variables $s_{ij}=1,0$ indicating respectively if the directed link  is matched or not.
Our goal is to find  the minimal set of variables $\{s_{ij}\}$ that satisfy the following  condition of matching, 
\bea
\sum_{j\in \partial_{+}i}s_{ij}\leq 1,\ \ \ \ \ \ \sum_{j\in \partial_{-}i}s_{ji}\leq 1,
\eea 
where $\partial_- i$ indicates the set of  nodes $j$ that point to node $i$ in the directed network,  and $\partial_+ i$ indicates the set of nodes $j$ that  are pointed by node $i$.
If these constraints are satisfied each node $i$ of the network has at most one in-coming link that is matched, (i.e. one neighbour $j\in \partial_-i$ such that $s_{ji}=1$) and at most one outgoing link (one neighbour $j\in \partial_+i$ such that $s_{ij}=1$) that is matched.
The maximum matching problem can be cast on a statistical mechanics problem where we consider the energy 
\bea
E&=&2\sum_{i=1}^N \left(1-\sum_{j\in \partial_{-}i}s_{ji} \right)\nonumber \\
&=&\sum_{i=1}^N \left(1-\sum_{j\in \partial_{-}i}s_{ji} \right)+\sum_{i=1}^N \left(1-\sum_{j\in \partial_{+}i}s_{ij} \right)\nonumber \\
&=&2N_D
\label{energys}
\eea
with $N_D$ being the number of unmatched nodes in the network. We aim at finding the distribution $P(\{s_{ij}\})$ given by 
\bea
P(\{s_{ij}\})&=&\frac{e^{-\beta E}}{Z}\prod_{i=1}^N\theta\left(1-\sum_{j\in \partial_{+}i}s_{ij}\right)\nonumber\\ &&\times\prod_{i=1}^N\theta\left(1-\sum_{j\in \partial_{-}i}s_{ji}\right)
\label{psij}
\eea
where $\theta(x)=1$ for $x\geq 0$ and $\theta(x)=0$ for $x<0$ and where $Z$ is the normalization constant, that corresponds to the partition function of the statistical mechanics problem. In particular our aim is to find this distribution in the limit $\beta\to \infty$ in order to characterize the optimal (i.e. the maximum-sized) matching in the network.
The free-energy density of the problem $f(\beta)$ is defined as 
\bea
\beta Nf(\beta)=-\ln Z,
\eea
and the energy of the problem is therefore given by 
\bea
E=\frac{\partial [\beta Nf(\beta)]}{\partial \beta}.
\label{der}
\eea

\subsection{The BP equations }
The distribution $P(\{s_{ij}\})$ on a locally tree-like network   can be solved by  the BP message passing method by finding the messages that nearby nodes sent to each other.
 In particular we distinguish between messages going in the direction of the link,  $P_{i\to j}(s_{ij})$, and messages going in the opposite direction of the link,  $\hat{P}_{i\to j}(s_{ji})$. The BP equations for these messages are 
 \bea
 P_{i\to j}(s_{ij})&=&\frac{1}{{\cal D}_{i \to j}} \sum_{s_{ \{ik \} } \setminus s_{ij}, k \in \partial_{+}i}  \theta\left(1-\sum_{k\in \partial_{+}i}s_{ik}\right)\nonumber \\
&&\times \exp{ \left [-\beta \left(1- \sum_{k\in \partial_{+}i}s_{ik} \right) \right]}\nonumber \\
&&\times\prod_{k\in \partial_+i\setminus j}\hat{P}_{k\to i}(s_{ik}),\nonumber \\
 \hat{P}_{i\to j}(s_{ji})&=&\frac{1}{{\hat{\cal D}}_{i \to j}}   \sum_{s_{ \{ki \} } \setminus s_{ji}, k \in \partial_{-}i}\theta\left(1-\sum_{k\in \partial_{-}i}s_{ki}\right)\nonumber \\
  &&\times\exp{ \left [-\beta \left(1- \sum_{k\in \partial_{-}i}s_{ki} \right)  \right]}\nonumber \\
  &&\times\prod_{k\in \partial_-i\setminus j}{P}_{k\to i}(s_{ki}),
  \label{BPe}
 \eea
 where ${\cal D}_{i \to j}$ and ${\hat{\cal D}}_{i \to j}$ are normalization constants.
 The messages $\{P_{i\to j}(s_{ij})$,  $\hat{P}_{i\to j}(s_{ji})\}$ can be parametrized by the cavity fields $h_{i\to j}$ and $\hat{h}_{i\to j}$ defined by 
\bea
&\begin{array}{lr}
P_{i\to j}(s_{ij})=\frac{e^{\beta h_{i\to j}s_{ij}}}{1+e^{\beta h_{i\to j}}}\ \ \ \ \ \ \  \hat{P}_{i\to j}(s_{ji})=\frac{e^{\beta\hat{ h}_{i\to j}s_{ji}}}{1+e^{\beta \hat{h}_{i\to j}}}.
& \end{array}
 \eea  
In terms of the cavity fields, Eqs.~\eqref{BPe} reduce to the following set of equations,
\bea
h_{i\to j}&=&-\frac{1}{\beta} \log\left(e^{-\beta}+\sum_{k \in\partial_{+}i\setminus j}e^{\beta \hat{h}_{k\to i}}\right),\nonumber \\
\hat{h}_{i\to j}&=&-\frac{1}{\beta}\log \left (e^{-\beta}+\sum_{k\in\partial_{-}i\setminus j}e^{\beta {h}_{k\to i}}\right).
\label{BPs}
\eea
that were first derived in \cite{Liu} for this problem. 

In the Bethe approximation, the probability distribution $P(\{s_{ij}\})$ is given by 
\bea
P_{Bethe}(\{s_{ij}\})=\prod_{i=1}^N P_i(\underline{S}_i)\left(\prod_{<i,j>}P_{ij}(s_{ij})\right)^{-1}
\label{uno}
\eea
where $P_i(\underline{S}_i)$ and $P_{ij}(s_{ij})$ are the marginal distribution over the nodes and the links of the network, that can be computed in terms of the 
cavity messages $P_{i\to j}(s_{ij})$,  $\hat{P}_{i\to j}(s_{ji})$, or equivalently the cavity fields $h_{i\to j}$ and $\hat{h}_{i\to j}$. They read
\bea
\label{due}
P_i(\underline{S}_i)&=&\frac{e^{-\beta[(1-\sum_{k\in \partial_{+}i}s_{ik})+(1-\sum_{k\in \partial_{-}i}s_{ki})]}}{{\cal C}_i}\nonumber\\ \\&&\times\theta\left(1-\sum_{k\in \partial_{+}i}s_{ik}\right)\theta\left(1-\sum_{k\in \partial_{-}i}s_{ki}\right)\nonumber \\&&\times\prod_{k\in \partial_+i}\hat{P}_{k\to i}(s_{ik})\prod_{k\in \partial_-i}{P}_{k\to i}(s_{ki})\nonumber \\
\label{tre}
P_{ij}(s_{ij})&=&\frac{1}{{\cal C}_{ij}}{P}_{i\to j}(s_{ij})\hat{P}_{j\to i}(s_{ij})
\eea 
where ${\cal C}_i$ and ${\cal C}_{ij}$ are normalization constant given by 
\bea
\label{cost1}
{\cal C}_i&=&\left(e^{-\beta}+\sum_{k\in\partial_+i}e^{\beta \hat{h}_{k\to i}}\right)\left(e^{-\beta}+\sum_{k\in\partial_-i}e^{\beta {h}_{k\to i}}\right) \nonumber\\
&&\times\prod_{k\in \partial_+i}\hat{P}_{k\to i}(0)\prod_{k\in \partial_-i}{P}_{k\to i}(0) \\
{\cal C}_{ij}&=&(1+e^{\beta (h_{i\to j}+\hat{h}_{j\to i})}){P}_{i\to j}(0)\hat{P}_{j\to i}(0).
\label{cost2}
\eea
\subsection{Free energy and energy of the problem}
The free energy of the problem can be found by evaluating the Gibbs free energy $F_{Gibbs}$ given by 
\bea
\beta F_{Gibbs}=\sum_{\{s_{ij}\}}P(\{s_{ij}\})\log\left(\frac{P(\{s_{ij}\})}{e^{-\beta E}\psi(\{s_{ij}\})}\right)
\eea
for  $P(\{s_{ij}\})=e^{-\beta E}\psi(\{s_{ij}\})/Z$, where $\psi(\{s_{ij}\})$  indicates the constraints
\bea
\psi(\{s_{ij}\})=\prod_{i=1}^N\left[ \theta\left(1-\sum_{j\in \partial_{+}i}s_{ij}\right) \theta\left(1-\sum_{j\in \partial_{-}i}s_{ji}\right)\right].
\eea
The distribution $P(\{s_{ij}\})=e^{-\beta E}\psi(\{s_{ij}\})/Z$ can be computed in the Bethe approximation using  $(\ref{uno}),(\ref{due}), (\ref{tre})$ and the fixed-point solutions of the BP equations \eqref{BPe}.
The Gibbs free energy $F_{Gibbs}$ is minimal when calculated over  the probability distribution $P(\{s_{ij}\})$ given by Eq. $(\ref{uno})$ and indeed for this distribution we  have $\beta F_{Gibbs}=-\ln Z$.
From the previous equations we can approximate the Gibbs free energy as
\begin{equation}\label{betheF}
\beta F_{Bethe}=\sum_{<i,j>}\log ({\cal C}_{ij}) - \sum_{i=1}^N \log({\cal C}_i).
\end{equation}

Inserting Eqs.\eqref{cost1},\eqref{cost2} into \eqref{betheF}, we obtain  the free energy of this matching problem, given by \cite{Liu} i.e.
\bea
\beta Nf(\beta)&=&-\sum_{i=1}^N\left(e^{-\beta}+\sum_{k \in \partial_+i}e^{\beta \hat{h}_{k\to i}}\right)\nonumber \\&&-\sum_{i=1}^N\left(e^{-\beta}+\sum_{k \in \partial_-i}e^{\beta {h}_{k\to i}}\right)\nonumber \\
&&+\sum_{<i,j>}\ln \left(1+e^{\beta (h_{i\to j}+\hat{h}_{j\to i})}\right).
\eea
Using Eq.\eqref{der} we get the energy 
\bea
E&=&\sum_{i=1}^N \left[\frac{e^{-\beta}-\sum_{k\in \partial_{+}i}\hat{h}_{k\to i}e^{\beta \hat{h}_{k\to i}}}{e^{-\beta}+\sum_{k\in \partial_{+}i}e^{\beta \hat{h}_{k\to i}}}\right]\nonumber \\
&&+\sum_{i=1}^N \left[\frac{e^{-\beta}-\sum_{k\in \partial_{-}i}{h}_{k\to i}e^{\beta {h}_{k\to i}}}{e^{-\beta}+\sum_{k\in \partial_{-}i}e^{\beta {h}_{k\to i}}}\right]\nonumber \\
\nonumber \\
&&+\sum_{<i,j>}\frac{(h_{i\to j}+\hat{h}_{j\to i})e^{\beta(h_{i\to j}+\hat{h}_{j\to i})}}{ 1+e^{\beta(h_{i\to j}+\hat{h}_{j\to i})}}.
\label{Ebs}
\eea
\subsection{The $\beta\to \infty$ limit}
In the $\beta\to \infty $ limit, the energy of a maximum matching can be written as follows 
\bea
\hspace*{-6mm}\nonumber E & = & - \sum_{i=1}^N \max\left[-1, \max_{k\in\partial_{+}i} \hat{h}_{k \to i}\right] - \sum_{j=1}^N \max\left[-1, \max_{k\in\partial_{-}i} h_{k \to i}\right] \\
  & & + \sum_{<i,j>}\max\left[0, h_{i \to j}+\hat{h}_{j \to i}\right]
\eea
in which for each directed link $(i,j)$ the cavity fields $\{h_{i \to j}, \hat{h}_{i \to j}\}$ satisfy the zero-temperature Belief Propagation equations, also known as Max-Sum (MS) equations,
\begin{subequations}\label{MS}
\bea
\label{MSa} h_{i \to j}&=&-\max\left[-1, \max_{k\in\partial_{+}i\setminus j} \hat{h}_{k \to i}\right], \\
\label{MSb} \hat{h}_{i \to j}&=&-\max\left[-1, \max_{k\in\partial_{-}i\setminus j} h_{k \to i}\right],
\eea
\end{subequations}
where in these equations when a node $i$ has only one outgoing link pointing to node $j$, i.e. $|\partial_+i|=1$ we assume $h_{i\to j}=1$; similarly, when node $i$ has only one incoming link coming from node $j$, i.e. $|\partial_-i|=1$ we assume $\hat{h}_{i \to j}=1$.
In the infinite size limit, the MS equations are closed for cavity fields with support either on $\{-1,1\}$ or on $\{-1,0,1\}$ \cite{Lenka,Liu, Altarelli}. 
When multiple solutions coexist, the dynamically stable solutions of minimum energy are the correct solutions of the maximum matching problem. 
\section{BP/MS Equations in an ensemble of random networks with given degree distribution}
In a random network with given in-degree distribution $P^{in}(k)$ and out-degree distribution $P^{out}(k)$  the fields $h$ and the fields $\hat{h}$ have distributions ${\cal P}(h)$ and $\hat{{\cal P}}(\hat{h})$ respectively. In the limit  $\beta\to \infty$ in which we look for the optimal matching we have  that these distributions can be written as a sum of three delta functions, i.e.
\bea
{\cal P}(h)&=&w_1\delta(h-1)+w_2\delta(h+1)+w_3\delta(h)\nonumber \\
\hat{\cal P}(\hat{h})&=&\hat{w}_1\delta(\hat{h}-1)+\hat{w}_2\delta(\hat{h}+1)+\hat{w}_3\delta(\hat{h}),
\eea
where the variables $\{w_1,w_2,w_3\}$ and the variables $\{\hat{w}_1,\hat{w}_2,\hat{w}_3\}$ must satisfy the following normalization conditions, 
$w_1+w_2+w_3=1$ and $\hat{w}_1+\hat{w}_2+\hat{w}_3=1$.
The MS equations \eqref{MS} can be written as equations for the set of probabilities $\{w\},\{\hat{w}\}$ obtaining 
\bea
w_1&=&\sum_{k} \frac{k}{\avg{k}_{out}}P^{out}(k)(\hat{w}_2)^{k-1}\nonumber \\
w_2&=&\sum_{k}\frac{k}{\avg{k}_{out}}P^{out}(k)\left[1-(1-\hat{w}_1)^{k-1}\right]\nonumber \\
\hat{w}_1&=&\sum_{k} \frac{k}{\avg{k}_{in}}P^{in}(k) ({w}_2)^{k-1}\nonumber \\ 
\hat{w}_2&=&\sum_{k} \frac{k}{\avg{k}_{in}}P^{in}(k)\left[1-(1-{w}_1)^{k-1}\right],
\label{BP_infs}
\eea
with $w_3=1-w_1-w_2$ and $\hat{w}_3=1-\hat{w}_1-\hat{w}_2$.
Moreover, the energy  given by Eq.~$(\ref{Ebs})$ in the $\beta\to \infty $ can be expressed in terms of the distributions $\{w_i\}$ and $\{\hat{w}_i\}$ obtaining, 
\bea
\frac{E}{N}&=&\sum_{k}P^{out}(k)\left\{\left(\hat{w}_2\right)^{k}-\left[1-(1-\hat{w}_1)^{k}\right]\right\}\nonumber \\
&&\sum_{k}P^{in}(k)\left\{\left({w}_2\right)^{k}-\left[1-(1-{w}_1)^{k}\right]\right\}\nonumber \\
&&+{\avg{k}_{in}}\left[\hat{w}_{1}(1-w_2)+w_{1}(1-\hat{w}_2)\right].
\eea
In other words, the fraction of driver nodes $n_D=E/(2N)$ in the network can be simply expressed in terms of the distributions  $\{w_i\}$ and $\{\hat{w}_i\}$.
Eqs.~\eqref{BP_infs} can have multiple solutions for the variables  $\{w_i\}$ and $\{\hat{w}_i\}$. In order to select the correct solution of the matching problem one should ensure that the following three conditions are satisfied. \\
{\it i) The sets $\{w_i\}$ and $\{\hat{w}_i\}$ must indicate two probability distributions;} \\
{\it ii) The solution should be stable:}\ 
The solution of the system of Eqs.~$(\ref{BP_infs})$ should be stable under small perturbation of the values of the distributions $\{w_i\}$ and $\{\hat{w}_i\}$. We will consider the stability condition in detail in the following subsection.\\
{\it iii) Find the optimal stable solution:}\
If the system of Eqs.~$(\ref{BP_infs})$ has more than one solution that satisfies both conditions {\it i)} and {\it ii)}, in order to find the optimal matching one should select the solution with lowest energy $E$.

\subsection{Stability condition}
Here we consider  the stability of the replica-symmetric solution of Eqs.~\eqref{BP_infs} (see e.g. \cite{Ricci,Rivoire,Castellani,Lucibello} for discussions on the RS stability). The replica symmetry assumes that all cavity fields have the same distributions ${\cal P}(h)$ and $\hat{\cal P}(\hat{h})$, that  in the zero temperature limit can be parametrized by mixtures of delta functions. If we relax such assumption,  we have to enlarge the functional space by considering distributions ${\cal Q}[{\cal P}]$ and $\hat{\cal Q}[\hat{\cal P}]$ of cavity field distributions. There are two ways in which the replica-symmetric solution can be recovered in this enlarged functional space: 1)  ${\cal Q}[{\cal P}] = \delta[{\cal P}-{\cal P}^*]$ with ${\cal P}^*(h) = \sum_{\alpha}w_{\alpha}\delta(h-h_{\alpha})$, and 2) ${\cal Q}[{\cal P}] = \sum_{\alpha}w_{\alpha} \delta[{\cal P}-\delta(h-h_{\alpha})]$.  

In the first case, the replica symmetric solution can become unstable towards a functional ${\cal Q}$ with non-zero variance and this corresponds to the dynamical instability of the solutions under iteration of the Eqs.~\eqref{BP_infs}. In other words, the instability means that the distribution of cavity fields does not actually concentrate around discrete values, therefore the corresponding solution is not reachable from any finite temperature. In order to evaluate this type of instability we compute the Jacobian of the system of Eqs.~\eqref{BP_infs} and impose that all its eigenvalues have modulus less than one. The $6\times6$ Jacobian matrix reads  
\begin{widetext}
\begin{equation}
J=\left(\begin{array}{cccccc}0&0&0&0&G_{1,out}^{\prime}(\hat{w}_2)&0\\ 0&0&0&G_{1,out}^{\prime}(1-\hat{w}_1)&0&0\\ -1&-1&0&0&0 &0\\
0& G_{1,in}^{\prime}(w_2)&0 &0& 0&0\\G_{1,in}^{\prime}(1-w_1)&0&0&0 &0&0\\ 0&0&0&-1&-1&0\end{array}\right).
\end{equation}
\end{widetext}
where 
\bea
G_{1,in}(x)&=&\sum_{k}\frac{k}{\avg{k}_{in}}P^{in}(k)x^{k-1}\nonumber \\
G_{1,in}^{\prime}(x)&=&\sum_{k}\frac{k(k-1)}{\avg{k}_{in}}P^{in}(k)x^{k-2}\nonumber \\
G_{1,out}(x)&=&\sum_{k}\frac{k}{\avg{k}_{out}}P^{out}(k)x^{k-1}\nonumber \\
G_{1,out}^{\prime}(x)&=&\sum_{k}\frac{k(k-1)}{\avg{k}_{out}}P^{out}(k)x^{k-2},
\eea
with $\avg{k}_{in}=\avg{k}_{out}$.
Two eigenvalues are zero, the other four have degenerate modulus, therefore the stability conditions are
\bea\label{conditions}
G_{1,in}^{\prime}(1-w_1)G_{1,out}^{\prime}(\hat{w_2})&<&1,
\nonumber \\
G_{1,out}^{\prime}(1-\hat{w}_1)G_{1,in}^{\prime}({w_2})&<&1.
\eea

In the second case, we have to consider a different type of instability  (called {\em bug proliferation}) that occurs because of a discrete change in the distribution that propagates through the network. We compute the probability $T(h_{\alpha} \to h_{\alpha'}| \hat{h}_{\beta} \to \hat{h}_{\beta'})$ that a certain node has a set of incoming fields such that it causes a cavity field $h_{\alpha}$ to change into $h_{\alpha'}$ as a consequence of the fact that one of its $k-1$ parents nodes changed from $\hat{h}_{\beta}$ to $\hat{h}_{\beta'}$. This gives,
\begin{widetext}
\begin{eqnarray}
\nonumber &&T(1\to -1|-1 \to 1) =  T(-1\to 1|1 \to -1)   =\hat{w}_2^{k-2}\\
 \nonumber &&T(1\to 0|-1 \to 0)  =  T(0\to 1|0 \to -1)  =  \hat{w}_2^{k-2}\\
\nonumber &&T(-1\to 0|1 \to 0)  =  T(0\to -1|0 \to 1)  =  (1-\hat{w}_1)^{k-2}\\
\nonumber &&T(-1\to 0|1 \to -1) =  T(0\to -1|-1 \to 1) =  (1-\hat{w}_1)^{k-2}-\hat{w}_2^{k-2}.
\end{eqnarray}
We have similar equations for the other set of cavity fields by replacing $\{w_1,w_2,w_3\}$ with $\{\hat{w}_1,\hat{w}_2,\hat{w}_3\}$. Consider one of these events, the probability that the out-coming (respectively in-coming) link in which a change occurs belongs to a node of degree $k$ is $k P^{out}(k)/\avg{k}_{out}$ (respectively $k P^{in}(k)/\avg{k}_{in}$) and this change affects $k-1$ other messages. When we average the possible perturbations for the $h$ fields and the $\hat{h}$ fields over the degree distributions, we get a $12\times 12$ block matrix  $\left(\begin{array}{c|c}  0 & T \\ \hline \hat{T} & 0\end{array}\right)$ with 
\begin{equation}
T=\left(\begin{array}{cccccccccccc}
 0   & 0 & 0   & 0 & 0 & G_{1,out}^{\prime}(\hat{w}_2)  \\
 0   & 0 & G_{1,out}^{\prime}(1-\hat{w}_1)-G_{1,out}^{\prime}(\hat{w}_2) & 0 & G_{1,out}^{\prime}(\hat{w}_2) & 0  \\
 0   & 0 & 0   & G_{1,out}^{\prime}(\hat{w}_2) & 0 & 0  \\
 0   & 0 & G_{1,out}^{\prime}(1-\hat{w}_1)   & 0 & 0 & 0  \\
 G_{1,out}^{\prime}(1-\hat{w}_1)-G_{1,out}^{\prime}(\hat{w}_2) & G_{1,out}^{\prime}(\hat{w}_2) & 0   & 0 & 0 & 0  \\
 G_{1,out}^{\prime}(1-\hat{w}_1)   & 0 & 0   & 0 & 0 & 0 
\end{array}\right)
\end{equation}
\begin{equation}
\hat{T}=\left(\begin{array}{cccccccccccc}
 0   & 0 & 0   & 0 & 0 & G_{1,in}^{\prime}(w_2)  \\
 0   & 0 & G_{1,in}^{\prime}(1-w_1)-G_{1,in}^{\prime}(w_2) & 0 & G_{1,in}^{\prime}(w_2) & 0  \\
 0   & 0 & 0   & G_{1,in}^{\prime}(w_2) & 0 & 0  \\
 0   & 0 & G_{1,in}^{\prime}(1-w_1)   & 0 & 0 & 0  \\
 G_{1,in}^{\prime}(1-w_1)-G_{1,in}^{\prime}(w_2) & G_{1,in}^{\prime}(w_2) & 0   & 0 & 0 & 0  \\
 G_{1,in}^{\prime}(1-w_1)   & 0 & 0   & 0 & 0 & 0 
\end{array}\right).
\end{equation}
\end{widetext}
Calculating the eigenvalues of the matrix, and imposing that their modulus is less than one, we obtain the following  stability conditions 
\bea
G_{1,in}^{\prime}(1-w_1)G_{1,out}^{\prime}(\hat{w}_2) &  < & 1,\nonumber\\
G_{1,out}^{\prime}(1-\hat{w}_1)G_{1,in}^{\prime}(w_2) &  < & 1,\nonumber \\
G_{1,in}^{\prime}(w_2)G_{1,out}^{\prime}(\hat{w}_2) &  < &  1.
\label{conditions-due}
\eea 

As a consequence of the normalization conditions on the $\{w_i\}_{i=1,2,3}$ and on the $\{\hat{w}_i\}_{i=1,2,3}$ we have $1-w_1\geq w_2$ and similarly $1-\hat{w}_1>\hat{w}_2$, therefore the last equation of Eqs.~\eqref{conditions-due} is redundant and therefore the stability conditions for this case are  the same as in Eqs.~\eqref{conditions}, i.e.
\bea
G_{1,in}^{\prime}(1-w_1)G_{1,out}^{\prime}(\hat{w}_2) &  < & 1,\nonumber\\
G_{1,out}^{\prime}(1-\hat{w}_1)G_{1,in}^{\prime}(w_2) & < & 1.
\label{conditions-tre}
\eea

By considering the zero-energy solution $w_1=w_2=\hat{w}_1=\hat{w}_2=0$ and $w_3=\hat{w}_3=1$, emerging for $P^{in}(1)=P^{out}(1)=0$, both stability criteria imply the condition in
Eq.~$(6)$ of the main text that we rewrite here for convenience, 
\bea
\hspace*{-3mm}P^{out}(2)<\frac{{\avg{k}_{in}}^2}{2\avg{k(k-1)}_{in}}, \ P^{in}(2)<\frac{{\avg{k}_{in}}^2}{2\avg{k(k-1)}_{out}}.
\label{stabilitys}
\eea
Notice that for $P^{in}(1)=P^{out}(1)=0$ there is also the zero energy solution  
$w_1=0, w_2=1,\hat{w}_1=1,\hat{w}_2=0$ and the symmetric solution
$w_1=1, w_2=0,\hat{w}_1=0,\hat{w}_2=1$. The first solution is stable when the stability  conditions given by Eqs.~\eqref{conditions} are satisfied, i.e. when  
\bea
G'_{1,in}(1)G'_{1,out}(0)&=&\frac{\avg{k(k-1)}_{in}}{\avg{k}_{in}}\frac{2P^{out}(2)}{\avg{k}_{out}}<1,
\eea
the second solution is stable when the following condition is satisfied 
\bea
G'_{1,in}(0)G'_{1,out}(1)&=&\frac{\avg{k(k-1)}_{out}}{\avg{k}_{out}}\frac{2P^{in}(2)}{\avg{k}_{in}}<1.
\eea
Therefore, when $P^{in}(k)=P^{out}(k)$, these solutions are stable under the same conditions in which the solution  $w_1=w_2=\hat{w}_1=\hat{w}_2=0$ is stable, and all these solutions correspond to the same value of the energy density $E/N=0$.  

\section{Number of driver nodes }
\begin{figure}[t]
\begin{center}
\includegraphics[width=0.9\columnwidth]{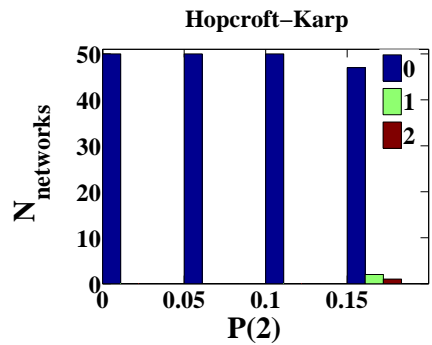} 
\end{center}
\caption{Histograms showing the number of network realizations that, out of a total of 50 realizations, show a certain number of driver nodes $N_D$ in the region of phase space in  which BP predicts zero fraction of driver nodes $n_D=0$. The different histograms are displayed as a function of $P(2)$ for in-degree and out-degree distributions as in Eq.~(7) of the main text with $P(1)=0$ and $\gamma=2.3$. The size of the  networks  is of $N=10^4$. The histogram refers to the exact matching algorithm by Hopcroft and Karp \cite{Hopcroft}. As long as we are far from the stability conditions $P(2)= 0.181947$, these results show that the expected number of driver nodes is consistent with $N_D=0$.} 
\label{fig:ND}
\end{figure}

The BP equations solving the maximum matching problem on a random network ensemble are expected to give the correct value for density of driver nodes in the limit of large networks $N\to \infty$. In particular, in the region in which BP predicts a zero fraction of driver nodes $n_D$, the BP algorithm does not guarantee that the exact number of driver nodes is zero, i.e.  $N_D=0$. Nevertheless in our simulations, by running the Hopcroft-Karp algorithm \cite{Hopcroft} on finite networks in the region where BP predicts a zero fraction of driver nodes, i.e. $n_D=0$, we have always found that, as soon as we are sufficiently far from the boundary of the region defined by the stability conditions, the networks have a number of driver nodes equal to zero, i.e. $N_D=0$.
In Fig.~$\ref{fig:ND}$ we show the histogram of the results obtained by the Hopcroft-Karp algorithm corresponding to the points of Fig. 3 of the main text with predicted zero fraction, i.e. $n_D=0$ of driver nodes.

\section{Improving the controllability of scale-free networks}
\begin{figure}[t]
\begin{center}
\includegraphics[width=0.8\columnwidth]{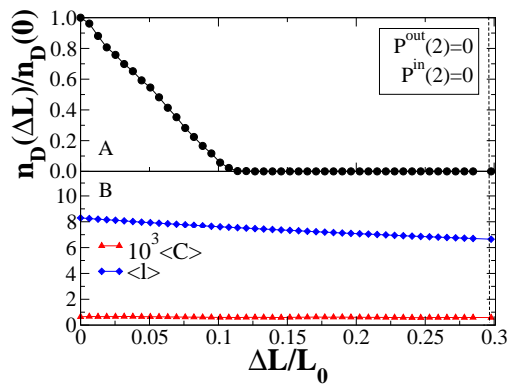} 
\end{center}
\caption{Fraction of driver nodes $n_D(\Delta L)/n_D(0)$ (panel A) average clustering coefficient $\avg{C}$ and average distance $\avg{l}$ (panel B) of the network as a function of the fraction of added links to low degree nodes. The results are obtained solving the MS equations. The initial network is  a power-law network with in-degree distribution equal to out-degree distribution,  $N=10^4$ nodes, and    power-law exponent $\gamma=3$. The symbol $\Delta L$ indicates the number of added links to the network, whereas $L_0$ indicates the initial number of links of the network. 
} 
\label{fig:gamma3}
\end{figure}

In the section {\it Improving the controllability of a network} of the main text we gave an example of a power-law network with in-degree distribution equal to out-degree distribution,  $N=10^4$ nodes, and   power-law exponent $\gamma=2.3$. We showed that in this particular case our recipe was quite demanding in terms of fraction of links needed to reach the full controllability of the network. Nevertheless, if we keep the same initial average degree and we consider the degree distributions with a power-law exponent $\gamma=3$, implying that we start from a minimum in-degree and our-degree equal to $2$,  the fraction of links for the full controllability drops to $13\%$ (see Fig.  $\ref{fig:gamma3}$).


\begin{figure}[!htb]
\begin{center}
\includegraphics[width=0.9\columnwidth]{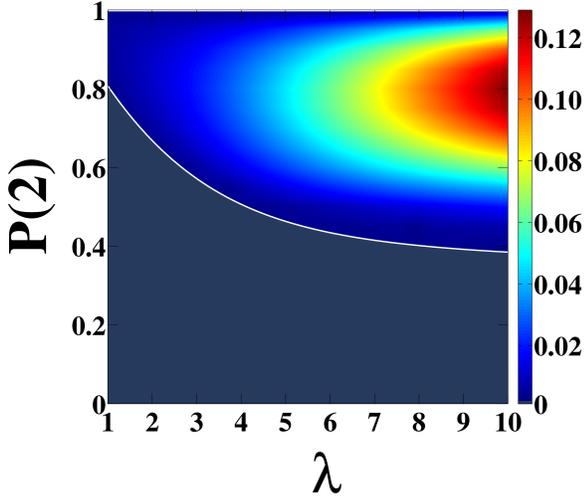} 
\end{center}
\caption{Phase diagram indicating the density of driver nodes $n_D$ (indicated according to the color code on the left) as a function of the parameters $\lambda$ and $P(2)$ for networks of nodes with degree distribution given by Eq. $(\ref{pkinoutpoisson})$ and $P(1)=0$. The density of driver nodes is obtained by numerically solving Eqs. $(\ref{BP_infs})$. The solid line indicates the stability line.} 
\label{fig:Poissonphaseplot}
\end{figure}
\section{Poisson networks}
In the main text of the paper we have assessed the role of low-degree nodes in the controllability of networks,  especially considering uncorrelated random graphs with power-law degree distribution. We consider now Poisson networks with the following degree distribution
\bea
P^{in}(k)=P^{out}(k)=\left\{\begin{array}{lcc}
P(1) &\mbox{if}&{k=1}\\
P(2)&\mbox{if}& k=2 \\
C\frac{\lambda^k}{k!} &\mbox{if} &k\in [3,\infty]\end{array}\right.
\label{pkinoutpoisson}
\eea
with $C$ a constant determined by normalization. We especially focus on the situation in which $P(1)=0$ and the stability condition for the solution $\{w_1,w_2,w_3\}=\{0,0,1\}$, $\{\hat{w}_1,\hat{w}_2,\hat{w}_3\}=\{0,0,1\}$ reads
\begin{equation}
P(2)\le\frac{\Avg{k}^2}{2(\Avg{k^2}-\Avg{k})} 
\end{equation}
where $\Avg{k}$ and $\Avg{k^2}$ can be easily expressed as
\bea
\Avg{k}&=&2P(2)+(1-P(2)) \frac{\lambda(e^{\lambda}-1-\lambda)}{e^{\lambda}-1-\lambda-\lambda^2/2}\\
\Avg{k^2}&=&4P(2)+(1-P(2))\frac{e^{\lambda}(\lambda+\lambda^2)-\lambda-2\lambda^2}{e^{\lambda}-1-\lambda-\lambda^2/2}
\eea
In Fig.~$\ref{fig:Poissonphaseplot}$ we show the phase diagram pointing out the fraction of driver nodes $n_D$ as a function of the parameters $\lambda$ and $P(2)$. 
The dark grey area  defines the region where the zero-energy solution is stable, hence the network has an infinitesimal fraction of driver nodes ($n_D =0$). Outside this region, the minimum fraction of driver nodes necessary for a full network control is displayed (lowest stable solution of the MS equations).

\section{Improving the controllability of Poisson networks}
\begin{figure}[h]
\begin{center}
\includegraphics[width=0.9\columnwidth]{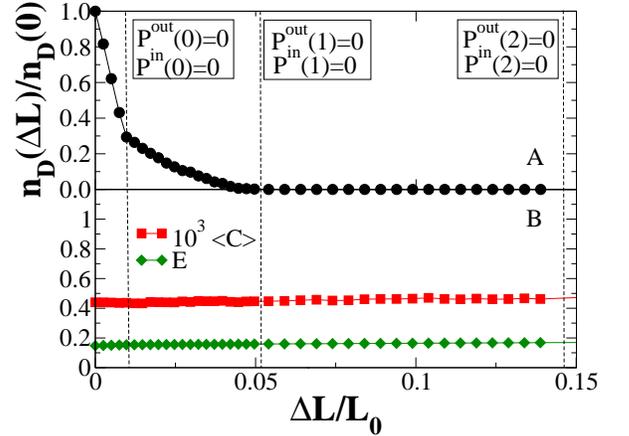} 
\end{center}
\caption{Fraction of driver nodes $n_D(\Delta L)/n_D(0)$(panel A) average clustering coefficient $\avg{C}$ and efficiency  $E$ (panel B) of the network as a function of the fraction of added links to low degree nodes. The results are obtained solving the MS equations with the Belief Propagation algorithm. The initial network is  a Poisson  network with in-degree distribution equal to out degree distribution,  $N=10^4$ nodes, and average degree $c=4$. The symbol $\Delta L$ indicates the number of added links to the network, whereas $L_0$ indicates the initial number of links of the network. The links are added to low degree nodes in the following way. First links are added to nodes of in-degree and out-degree $0$  and then links are added to nodes of in-degree and out-degree $1$ and then to nodes of in-degree and out-degree $2$ as described in the main text. This strategy can be used to increase the controllability of networks.
} 
\label{fig:Poisson}
\end{figure}

In the main text of the paper we have described an algorithm that can improve the controllability of networks by adding links to it and reducing the number of nodes with in-degree and out-degree smaller than 3. While in the main text we show that such algorithm can be used to improve the controllability of scale-free networks, here we show that the same algorithm can be used to improve the controllability  also of Poisson networks. In fact this approach can be applied to networks with any type of degree distribution.
In Figure~\ref{fig:Poisson} we display the fraction $n_D(\Delta L)$ of driver nodes when we add $\Delta L$ links in the network divided by its initial  value $n_D(0)$  where the network has a Poisson degree distribution and average degree $c=4$. We note that in this case the fraction of links that need to be added to have full controllability is of the order of $5\%$. Here we have chosen to display the efficiency $E$ instead of the average distance $\avg{l}$ because the network, specially at the beginning, is not fully connected.\\
When $P^{in}(1)=P^{out}(1)=0$ the displayed network has $P^{in}(2)=P^{out}(2)\approx 0.21$ and it becomes fully controllable.

\end{document}